\pdfoutput=1
\documentclass[doublecol]{epl2}
\usepackage{amsmath}
\usepackage{graphicx}
\usepackage{color}
\usepackage{hyperref}
\usepackage{psfrag}
\usepackage{gensymb}
\usepackage{enumerate}
\usepackage{stmaryrd}

\title{Relevance of visco-plastic theory in a multi-directional inhomogeneous granular flow}
\shorttitle{}

\author{P.-P. Cortet\inst{1,2} \and D. Bonamy\inst{2} \and F. Daviaud\inst{1} \and O. Dauchot\inst{1} \and B. Dubrulle\inst{1} \and M. Renouf\inst{3}}
\shortauthor{P.-P. Cortet \etal}

\institute{
  \inst{1} CEA, IRAMIS, SPEC, CNRS URA 2464, Grp. Instabilit\'es \& Turbulence, 91191 Gif-sur-Yvette, France\\
  \inst{2} CEA, IRAMIS, SPCSI, Grp. Complex Systems \& Fracture, 91191 Gif-sur-Yvette, France\\
  \inst{3} LaMCoS, CNRS UMR 5259, INSA Lyon, 18-20 rue des sciences, 69621 Villeurbanne, France}


\abstract{We confront a recent visco-plastic description of dense
granular flows [P. Jop \etal, Nature, {\bf 441} (2006) 727] with
multi-directional inhomogeneous steady flows observed in
non-smooth contact dynamics simulations of 2D half-filled rotating
drums. Special attention is paid to check separately the two
underlying fundamental statements into which the considered theory
can be recast, namely (i) a single relation between the invariants
of stress and strain rate tensors and (ii) the alignment between
these tensors. Interestingly, the first prediction is fairly well
verified over more than four decades of small strain rate, from
the surface rapid flow to the quasi-static creep phase, where it
is usually believed to fail because of jamming. On the other hand,
the alignment between stress and strain rate tensors is shown to
fail over the whole flow, what yields an apparent violation of the
visco-plastic rheology when applied without care. In the
quasi-static phase, the particularly large misalignment is
conjectured to be related to transient dilatancy effects.}

\begin{document}

\maketitle

Despite granular flows ubiquity in nature and industry, their
description still remains an open issue. Three regimes are usually
distinguished: The rapid dilute flow -- gaseous-like -- regime,
where grains interact through binary collisions, is described
within the framework of the kinetic theory \cite{Savage81_jfm};
The slow flow -- solid-like -- regime, where only friction between
grains is relevant, is described using soil mechanics
\cite{nedderman92_book}. In between, there is a dense flow --
liquid-like -- regime, where inertia becomes important, friction
remaining however relevant. This last regime has been widely
investigated experimentally and numerically (see
\cite{Gdrmidi04_epje} for a review) in various flow configurations
and has been the focus of many theoretical works
\cite{Savage98_jfm,Losert00_prl,Mills1999_epl,Aranson02_pre,Lemaitre02_prl}.
However, while these descriptions reproduce some of the features
observed experimentally, a unified description is still missing.
In particular, the possibility for a single and local constitutive
law within the framework of continuum mechanics is still debated.

Progresses have been made recently in this context. Dimensional
analysis \cite{Iordanoff04_asmejt,Gdrmidi04_epje,Dacruz05_pre} was
used to show that, in simple incompressible unidirectional
uniformly sheared flows, bead diameter $d$, grain density $\rho$,
global shear stress $\tau$, global pressure $P$ and global shear
rate $\dot\gamma$ should be related as $\tau/P=\mu(I)$, where
$\mu(I)$ is an effective friction coefficient, function of the
so-called \textit{Inertial Number} $I=\dot{\gamma}d/\sqrt{P/\rho}$
expected to control locally the rheology of the material. Such
relation has first been evidenced, numerically, in plane shear
\cite{Iordanoff04_asmejt,Dacruz05_pre} and, experimentally, in
inclined plane \cite{Gdrmidi04_epje} configurations. Local
tensorial extension of this relation was recently proposed by Jop
\textit{et al.} \cite{Jop06_nature} as a constitutive law for
dense granular flows:
\begin{eqnarray}\label{rheo_2D_bis}
\tau_{ij}=\mu(I)\,\frac{P}{|\dot{\gamma}^d|}\, \dot{\gamma}_{ij}^d
\quad \mathrm{with} \quad
I=\frac{|\dot{\gamma}^d|d}{\sqrt{P/\rho}}
\end{eqnarray}
where $\tau_{ij}=\sigma_{ij}+P\delta_{ij}$ denotes the deviatoric
part of the stress tensor, $\dot{\gamma}_{ij}=1/2\,(\partial_i
v_j+\partial_j v_i)$ is the strain rate tensor,
$|\dot{\gamma^d}|=\sqrt{\frac{1}{2}
\dot{\gamma}^d_{ij}\dot{\gamma}^d_{ij}}$ the norm of its
deviatoric part
$\dot{\gamma}_{ij}^d=\dot{\gamma}_{ij}-1/2\,\dot{\gamma}_{kk}\delta_{ij}$,
$P=-1/2\,\sigma_{kk}$ the local pressure and $v_i$ the components
of the velocity. In the compressible case, a consistent local
rheology would imply in addition a univocal dependance
$\nu=\nu(I)$ of the volume fraction on the inertial number
\cite{Dacruz05_pre}. Introducing equation (\ref{rheo_2D_bis}) in
the momentum equation, Jop \textit{et al.} succeeded to fit the
surface velocity profile for a steady unidirectional flow down an
inclined plane with walls. More recently, relation
(\ref{rheo_2D_bis}) was confronted with simulations of complex
multi-directional non-steady inhomogeneous flow -- the collapse of
a cylinder of granular matter onto a horizontal plane
\cite{Lacaze09} -- and was shown to hold fairly well within the
fast flowing region \textit{i.e.} for $I$ ranging from $10^{-2}$
to $0.8$. In this last work, relation (\ref{rheo_2D_bis}) has
however been verified only for positions where stress and strain
rate tensors had parallel principal directions, excluding $10\%$
of data from the test. Therefore, despite these various
experimental and numerical results, the relevance and limits of
the visco-plastic relation (\ref{rheo_2D_bis}) remains to be
investigated in granular flows involving small values of inertial
number $I$ -- in which the stress and strain rate tensors are
particularly prone to be non-aligned -- as \textit{e.g.}
avalanches flows where large scale non-local effects are expected
\cite{Bonamy02,Radjai03,Pouliquen04,Deboeuf}.

In the present Letter, we confront this visco-plastic description
of granular flows with steady surface flows observed in a 2D
rotating drum, using extensive numerical simulations providing
large statistics. The motivation for studying this specific flow
is double since, still remaining in a ``simple'' 2D geometry, (i)
it provides a general highly multi-directional inhomogeneous
situation and (ii) it offers a wide range of inertial number:
$10^{-6}<I<5\times 10^{-2}$. In such a flow, we report an apparent
failure of the constitutive relation (\ref{rheo_2D_bis}) over the
whole range of $I$, when applied in the ``natural'' frame of the
flow. To overcome this failure, we highlight the fact that
equation (\ref{rheo_2D_bis}) is actually composed of two
independent results, namely (a) a single relation between the
invariants of stress and strain rate tensors and (b) the alignment
between these tensors, that constitute its two underlying
fundamental hypothesis. The system of these two conditions
actually recasts equation (\ref{rheo_2D_bis}) in a more
fundamental formulation. Then, the norms of the stress and strain
rate tensors are found to be univocally related through the
invariant relation derived from equation (\ref{rheo_2D_bis}) over
the whole range of $I$, meaning condition (a) is verified over the
whole flow. Finally, the failure of the tensorial relation
(\ref{rheo_2D_bis}) is shown to result from a misalignment between
the stress and strain rate tensors -- meaning condition (b) fails
-- contrary to what is assumed in previous works. Possible origins
of this misalignment are discussed: In the deep ``solid'' part of
the flow, the total deformation experienced by the granular
packing during its lifetime is less than one bead diameter. This
induces transient dilatancy effects -- from a Lagrangian point of
view -- that could be at the origin of the persistence of the
observed large misalignment.

\begin{figure}
\centerline{\includegraphics[width=8.8cm]{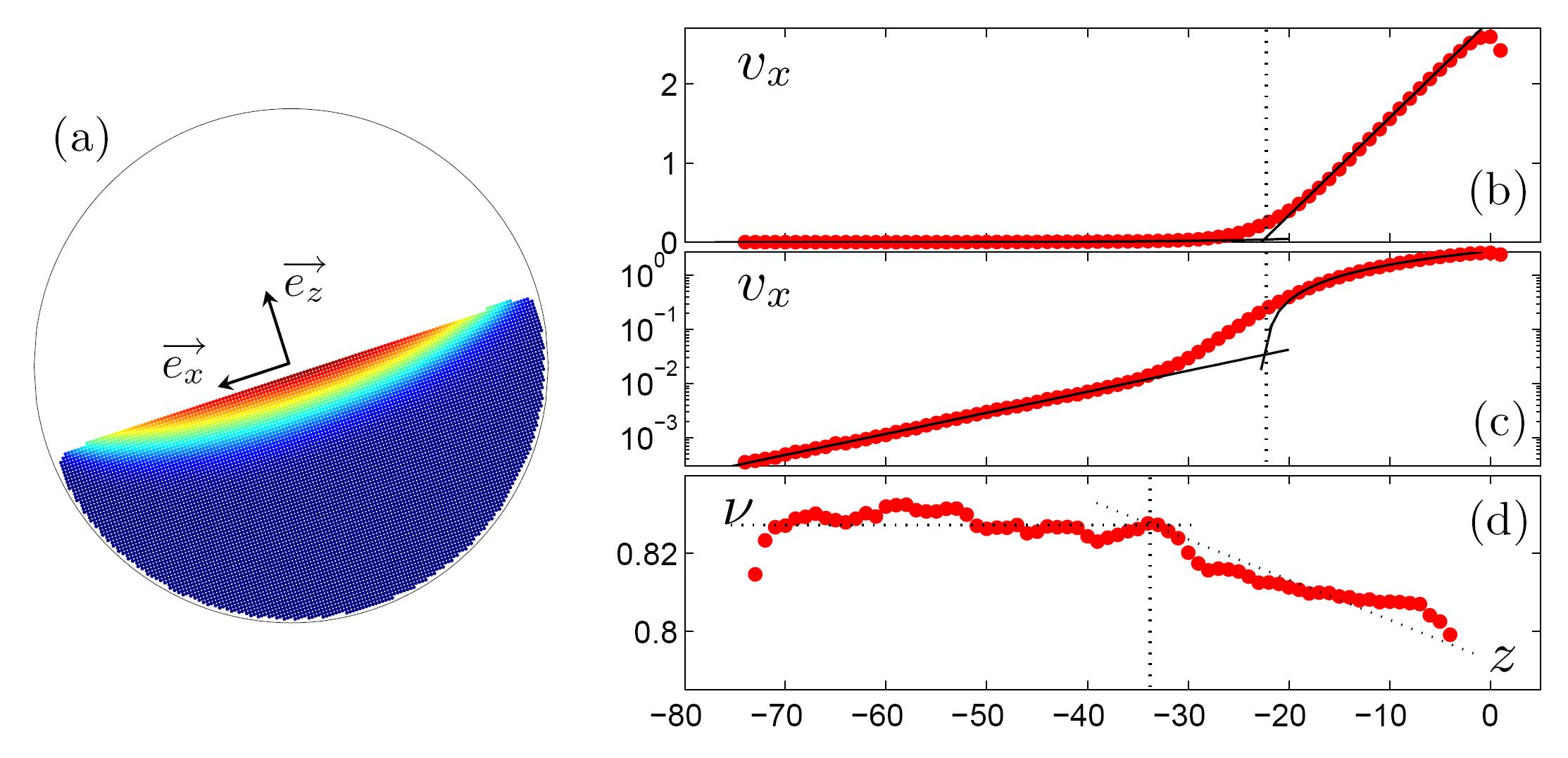}}
\caption{(a-c) time and ensemble averaged streamwise velocity
$v_x$ for a $\Omega=6$rpm rotation velocity: (a) map with
normalized colormap, (b) with linear and (c) logarithmic z-axis as
a function of the depth $z$ at the center of the drum ($x=0$). (d)
volume fraction $\nu$ as a function of the depth $z$ at the center
of the drum ($x=0$). Velocities are non-dimensionalized by
$\sqrt{gd}$ where $g$ is the gravity constant and $d$ the mean
bead diameter. In (b) and (c), solid lines are the result of a
linear and an exponential data fit.} \label{snapshot2D}
\end{figure}

The simulations have been performed using the Non-Smooth Contact
Dynamics method \cite{Moreau88_proc}. The scheme has been
described in detail in \cite{Renouf05_pof} and is briefly recalled
below. An immobile drum of diameter $D_0=45$cm is half-filled with
7183 rigid disks of density $\rho=2.7$g.cm$^{-3}$ and diameter
uniformly distributed between $3$ and $3.6$mm. The weak
polydispersity introduced in the packing prevents 2D ordering
effects. The normal restitution coefficient between two disks
(\textit{resp.} between disks and drum) is set to 0.46
(\textit{resp.} 0.46) and the friction coefficient to 0.4
(\textit{resp.} 0.95). Once the initial packing is stabilized, a
constant rotation velocity $\Omega$ ranging from $2$ to $12$ rpm
is imposed to the drum. After one round, a steady continuous
surface flow has developed. Then, one starts to capture 400
snapshots equally distributed over one rotation of the drum. For
each bead, one records the position $\overrightarrow{r}$ of its
center of mass and its instantaneous velocity $\overrightarrow{c}$
over the time step $\delta t = 6\times 10^{-3}$s. For each
rotation velocity, we performed $20$ experiments starting from
different initial packings. The reference frame $\Re$ is defined
as the frame rotating at $\Omega$ with the drum that co{\"i}ncides
with the reference frame
$\Re_0=(\overrightarrow{e_x},\overrightarrow{e_z})$ (cf. fig.
\ref{snapshot2D}), fixed in the laboratory, where
$\overrightarrow{e_x}$ (\textit{resp.} $\overrightarrow{e_z}$) is
parallel (\textit{resp.} perpendicular) to the mean flow free
surface. The mean values of the various Eulerian fields (volume
fraction, velocity and stress tensor) are obtained, at the grain
scale (see \cite{Renouf05_pof}) and without any further
coarse-graining, after both time and ensemble averaging over all
the 400 frames of all the 20 experiments.

Steady surface flows in rotating drums provide a continuum of flow
regimes from ``quasi-static' to ``liquid''. Figures
\ref{snapshot2D}(b) and (c) presents typical velocity profiles
$v_x(x=0,z)$ measured in our numerical simulations ``at the
center'' of the drum (\textit{i.e.} along the z-aligned slice at
$x=0$). Two phases can be identified: A flowing layer exhibiting a
linear velocity profile and a ``solid'' phase experiencing creep
motion where $v_x$ decays exponentially with $z$. Such velocity
profile corresponds well to experimental measurements in rotating
drum \cite{Gdrmidi04_epje,Rajchenbach00_ap,Bonamy02pof,Courrech}
as well as in heap geometry \cite{Komatsu01_prl,Gdrmidi04_epje}.
It is worth mentioning that the characteristic length $\xi$
associated with the exponential creep is found to be $\xi_{\rm}=12
\pm 2$ in our numerical drum over the range of explored rotating
velocity. This is significantly larger than that reported by
Crassous \textit{et al.} \cite{Crassous} in heap geometry, $\xi=1
\pm 0.2$. This difference may emphasize the role played by the
forcing on the selection of the velocity profiles within the bulk
of the material. Indeed, the forcing in rotating drum and heap are
very different since, from a Lagrangian viewpoint, the gravity is
rotating in the first case while it remains frozen in the second.
The volume fraction is found to be almost constant, around
$0.827\pm 0.05$, in the creeping layer while it decreases slightly
in the flowing layer, down to $0.80$ at $z=-4$ (cf. fig.
\ref{snapshot2D}(d)) as expected from dilatancy effects.

\begin{figure}
\includegraphics[width=8.8cm,clip]{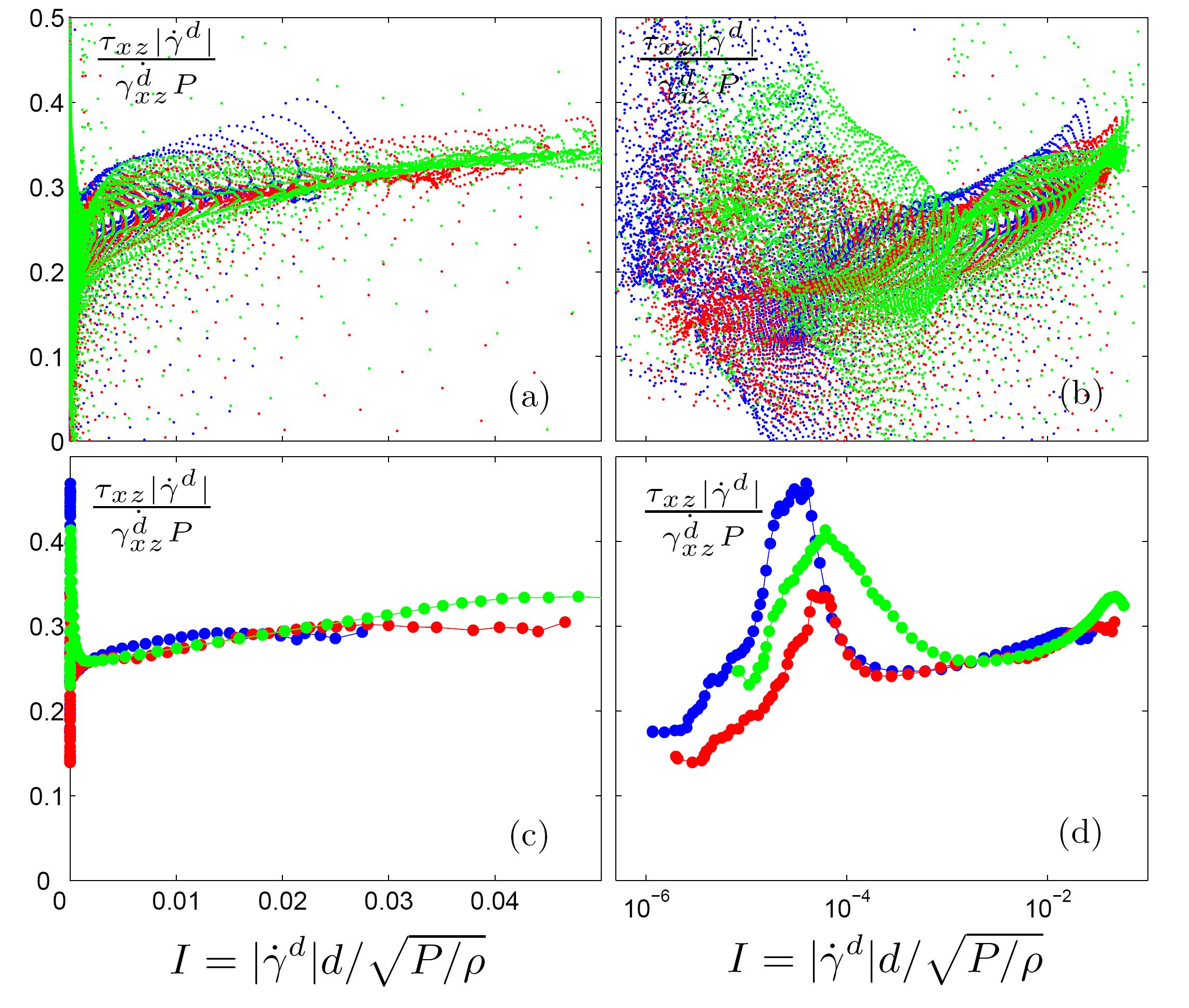}
\caption{Effective friction coefficient $\tau_{xz}
|\dot{\gamma}^d|/\dot{\gamma_{xz}^d} P$ as a function of the
inertial number for time and ensemble averaged data, (a) and (b)
over the whole 2D drum, (c) and (d) at the center of the drum ($x
= 0$), with linear, (a) and (c), and logarithmic, (b) and (d),
x-axis. Green/light gray points correspond to $\Omega=12$rpm,
red/intermediate gray points to $\Omega=6$rpm, and blue/dark gray
points to $\Omega=2$rpm.} \label{mu_I_nt}
\end{figure}

The full 2D fields of deviatoric stress $\tau_{ij}(x,z)$,
deviatoric strain rate $\dot{\gamma}^d_{ij}(x,z)$ and pressure
$P(x,z)$ are then computed to test the relevance of constitutive
equation (\ref{rheo_2D_bis}). Figures \ref{mu_I_nt}(a) and (b)
show the variations of the ratio $\tau_{xz}
|\dot{\gamma}^d|/\dot{\gamma_{xz}^d} P$ as a function of $I$ for
time and ensemble averaged data corresponding to the whole 2D drum
and for three different rotation velocities. If the rheology were
correct, these data would collapse on the single master curve
$\mu(I)$. This is clearly not the case as we observe an important
dispersion of the data -- even for the larger values of $I$.
Additionally, as one can see in figures \ref{mu_I_nt}(c) and (d),
the collapse predicted by equation (\ref{rheo_2D_bis}) even
does not work well (though better) when restricting data to the
slice $x=0$ located at the center of the drum, where the flow is
expected to be locally unidirectional ($\sslash$ to $\vec{e}_x$)
and almost homogeneous ($\partial_x\simeq0$) as in the plane shear
configuration. Moreover, the shape of these curves does not
correspond to the logarithmic shape previously evidenced in
\cite{Gdrmidi04_epje,Dacruz05_pre,Jop06_nature}, especially,
because of the unexpected bump of $\tau_{xz}
|\dot{\gamma}^d|/\dot{\gamma_{xz}^d} P$ around
$I=10^{-5}-10^{-4}$.

\begin{figure}
\includegraphics[width=8.8cm,clip]{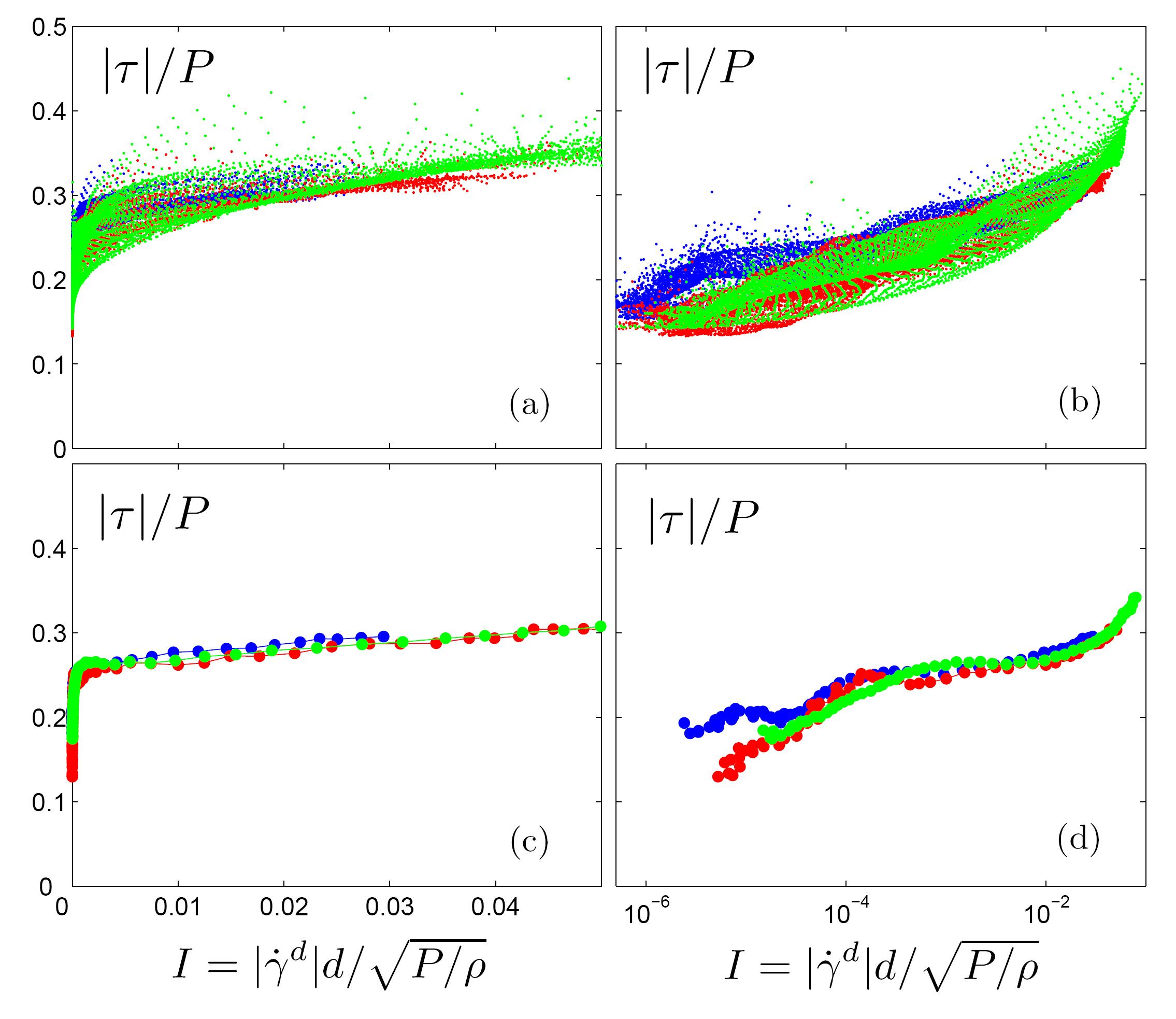}
\caption{Invariant effective friction coefficient $|\tau|/P$ as a
function of the inertial number for time and ensemble averaged
data, (a) and (b) over the whole 2D drum, (c) and (d) at the
center of the drum ($x = 0$), with linear, (a) and (c), and
logarithmic, (b) and (d), x-axis.  Same color/gray level code as
in figure \ref{mu_I_nt}.} \label{mu_I_norm}
\end{figure}

The multi-directionality of the considered flow implies that no
reference frame appears more natural than another to study
tensors. Therefore, it is reasonable to recast tensorial equations
into reference frame invariant relations. For granular flows,
searching for a rheology consists mainly in relating the
deviatoric stress tensor $\tau_{ij}$ to the deviatoric strain rate
tensor $\dot{\gamma}_{ij}^d$. Since both of these tensors are
symmetric with zero trace, both of them are fully characterized
through one invariant only, \textit{e.g.} their norm, and their
principal directions. Accordingly, the rheology of equation
(\ref{rheo_2D_bis}) can be recast in the following invariant way:
\begin{enumerate}[(a) ]
\item $\frac{|\tau|}{P}=\mu(I)$, \item $\tau_{ij}$ and
$\dot{\gamma}_{ij}^d$ have identical principal directions,
\end{enumerate}
where $|\tau|=\sqrt{\frac{1}{2}\tau_{ij}\tau_{ij}}$ is the
deviatoric stress tensor norm. In unidirectional flows, such as
those observed in plane shear geometry \cite{Dacruz05_pre} or down
to rough inclines \cite{Jop06_nature}, condition (b) is naturally
ensured by the flow symmetries so that verifying equation
(\ref{rheo_2D_bis}) reduces to verifying point (a) only. As we
will see thereafter, this is no more the case for a
multi-directional flow for which the alignment of the stress and
strain rate tensors cannot be assumed \cite{Ries,Depken}. The
split of equation (\ref{rheo_2D_bis}) into the two independent
conditions (a) and (b) recasts the rheology of Jop \textit{et al.}
into its two underlying fundamental assumptions which are
completely independent. In the sequel, we will see that the
invariant form of the visco-plastic rheology -- condition (a) --
actually holds well over the whole rotating drum and the whole
range of $I$ whatever condition (b) holds or not. This
reformulation of equation (\ref{rheo_2D_bis}) actually allows to
test the visco-plastic theory for very small values of $I$ down to
$10^{-5}$ where the condition on stress and strain rate tensor
alignment is largely violated.

Let us first consider point (a). In figure \ref{mu_I_norm}, we
have plotted the invariant effective friction coefficient
$|\tau|/P$ as a function of the inertial number for the same data
as in figure \ref{mu_I_nt}. We see that all the data appear to
collapse quite well on a single master curve $\mu(I)$
independently of the rotation velocity $\Omega$. The shape of this
master curve and the value of the corresponding friction
coefficient are perfectly consistent with the ones already
evidenced ($\mu=0.30\pm 0.10$ for $I<0.1$) in different granular
unidirectional flows
\cite{Gdrmidi04_epje,Dacruz05_pre,Jop06_nature}. It is important
to emphasize that this collapse is a non trivial result since we
have plotted data for the whole 2D highly inhomogeneous fields. In
figures \ref{mu_I_norm}(c) and (d), we restrict data to the slice
$x=0$ located at the center of the drum where the flow is expected
to be almost ``plane shear like''. We see that the $|\tau|/P$
profiles for the three considered rotation velocities fall with a
very good precision on the same master curve $\mu(I,x=0)$ for
values of $I$ as small as $10^{-5}$. Despite this last remark,
this collapse is a strong evidence for the possible relevance of
the constitutive law of equation (\ref{rheo_2D_bis}) in steady
inhomogeneous multi-directional flows. More importantly, it is the
first time such frictional rheology seems to hold over such a low
inertial number range ($10^{-5}<I<5\times 10^{-2}$) whereas it is
often expected to fail because of the jamming transition for
$I<10^{-3}$.

\begin{figure}[h]
\includegraphics[width=8.8cm]{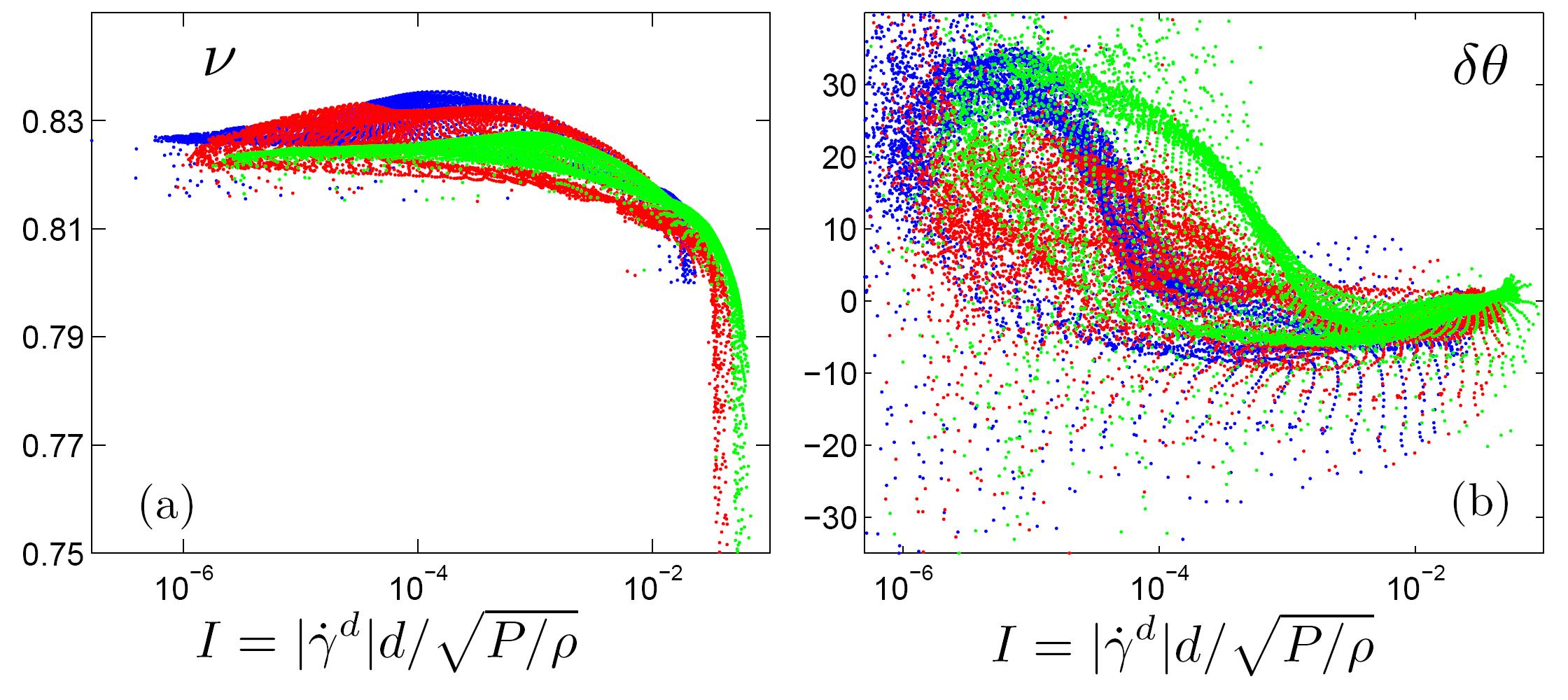}
\caption{(a) Volume fraction $\nu$ and (b) angle difference
between principal directions of $\tau_{ij}$ and
$\dot{\gamma}^d_{ij}$ as a function of the inertial number for
time and ensemble averaged data corresponding to the whole 2D
drum. Same color/gray level code as in figure \ref{mu_I_nt}.}
\label{nu_I}
\end{figure}

We already said that in compressive flows a local rheology would
imply a single dependence of the volume fraction $\nu$ with the
inertial number $I$. In figure \ref{nu_I}(a), we see that the data
corresponding to the whole 2D flow collapse indeed quite well on a
master curve $\nu=\nu(I)$ supporting partially the proposed
rheological approach.

\begin{figure}
\includegraphics[width=8.8cm]{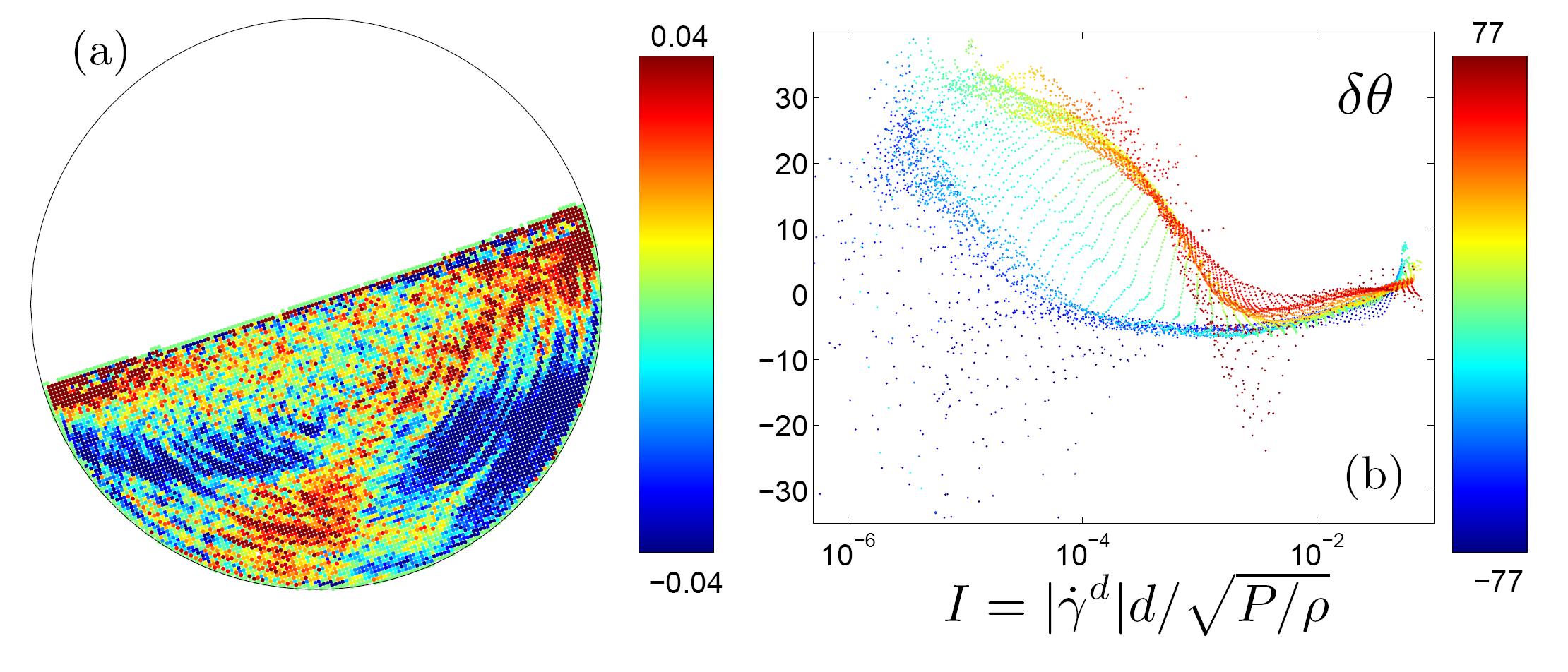}
\caption{(a) Map of the deviation $|\tau|/P-|\tau|/P(x=0,z)$ of
the effective friction coefficient $|\tau|/P$ from the one
$\mu(I,x=0)$ at the center of the drum for $\Omega=6$rpm. (b)
angle difference between principal directions of $\tau_{ij}$ and
$\dot{\gamma}^d_{ij}$ as a function of the inertial number for
$\Omega=12$rpm. In (b), data point colors are mapping the $x$
position in the drum.} \label{dev_map}
\end{figure}

The second step is now to check point (b) \textit{i.e.} the
collinearity between the deviatoric stress $\tau_{ij}$ and strain
rate $\dot{\gamma}_{ij}^d$ tensors. Figure \ref{nu_I}(b) shows the
angle difference $\delta \theta$ between the principal directions
of $\tau_{ij}$ and $\dot{\gamma}_{ij}^d$ as a function of the
inertial number. The collinearity is not verified since $\delta
\theta$ ranges from $\pm 10^\circ$, for the larger inertial
numbers ($5\times 10^{-4}<I<5\times 10^{-2}$) corresponding to the
``liquid'' surface flow, up to $35^\circ$ in the deeper part of
the flow, corresponding to smaller inertial numbers ($
10^{-6}<I<5\times 10^{-4}$). Such increase in the dispersion of
the tensors misalignment, as $I$ decreases, has also been reported
in shear zones \cite{Ries}. More importantly, there is no univocal
relation ruled by the inertial number such as $\delta
\theta=\delta \theta(I)$. Consequently, the fact the tensorial law
(\ref{rheo_2D_bis}) is not verified contrary to its invariant
version, relation (a), on the norms of the tensors holds is the
direct consequence of the misalignment of the stress and strain
rate tensors. Moreover, it shows, looking at the larger inertial
numbers in figures \ref{mu_I_nt} and \ref{nu_I}(b), that a
$10^\circ$ discrepancy between principal directions is already
significant to make the visco-plastic tensorial rheology, in its
non-invariant version of equation (\ref{rheo_2D_bis}), flatter.

\begin{figure}
\centerline{\includegraphics[width=8.8cm]{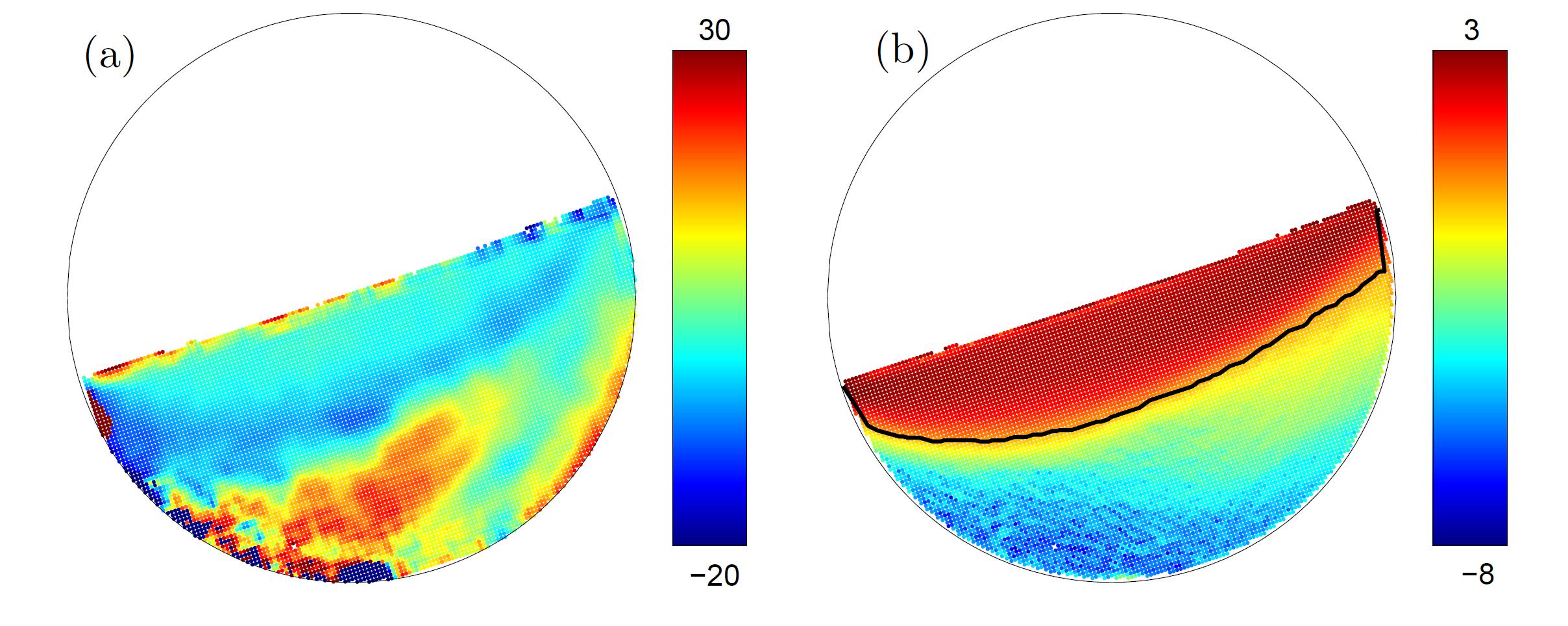}}
\caption{(a) Map of the angle difference between principal
directions of $\tau_{ij}$ and $\dot{\gamma}^d_{ij}$ for
$\Omega=12$rpm. (b) Map of the logarithm of the norm of the strain
rate tensor times the half drum rotation duration
$\log{|\dot{\gamma}^d|/2\Omega}$ for $\Omega=12$rpm. We indicate
with a black line the limit where this deformation
$\delta\equiv|\dot{\gamma}^d|/2\Omega=1$.} \label{gamma_theta}
\end{figure}

In figure \ref{mu_I_norm}, the collapse of the data on a master
curve $\mu(I)$ is not perfect and one may wonder if the dispersion
around the ``mean'' friction coefficient $\mu(I,x=0)$ measured at
the center of the drum is only due to a lack of statistics.
Therefore, in figure \ref{dev_map}(a), we plot the map of the
deviation of the effective friction coefficient from the one at
the center of the drum. One can observe a spatial structuration
with a quadripolar pattern. This spatial dependence of the profile
$|\tau|/P=\mu(I,x)$ with the position $x$ may be the consequence
of non-local effects reflecting the long range influence of the
boundaries through force-chains, but remains however an
unexplained feature. It is worth to notice that a pure geometrical
dependence of $|\tau|/P$ has been reported in \cite{Depken} where
the interest of studying the present regime of mixed geometrical
and shear rate dependence was highlighted. In parallel, in figures
\ref{dev_map}(b) and \ref{gamma_theta}(a), we see that the cloud
of data $\delta \theta$ \textit{vs.} $I$ shows as well a
non-trivial spatial structure. There is however no evidence for
any link between the spatial structures of the angle $\delta
\theta$ and of the deviation around $\mu(I,x=0)$, the origins of
which are still to be understood.

Previous studies \cite{Ries,Depken} have shown that the stress and
the strain rate tensor principal directions can differ
significantly due to transient dilatancy effects as long as the
initial conditions of the packing have not been forgotten,
\textit{i.e.} as long as the local deformation is less than $\sim
1$. Even if our flow is stationary, the transient effect reported
above could be the explanation for the large misalignment observed
in the deep part of the rotating drum (cf. fig.
\ref{gamma_theta}(a)). Indeed, in the ``solid'' part of the
granular matter, a given ``Lagrangian'' packing is created at the
bottom of the surface avalanche. Then, it has a short lifetime
reduced to half a drum rotation period $1/2\Omega$ corresponding
to the time needed for him to reach the top of the avalanche where
it is destructed. During this process, the deformation experienced
by the packing is quite small: in figure \ref{gamma_theta}(b), we
plot the map of the logarithm of the norm of the strain rate
tensor times the half drum rotation duration
$\log{|\dot{\gamma}^d|/2\Omega}$ as well as a black line
delineating where the deformation
$\delta=|\dot{\gamma}^d|/2\Omega$ equals unity. The ``solid''
granular matter under this line is experiencing a transient
deformation since it is less than one. Finally, we see that this
``transient'' zone matches quite well the zone (cf. fig.
\ref{gamma_theta}(a)) where $|\delta \theta|$ is larger than
$10^\circ$ making transient dilatancy effects a relevant
explanation for the persistence of large alignment discrepancies.
This observation does not imply that the spatial structure
observed for the angle $\delta \theta$ is due to transient
effects. On the contrary, this structure is created by the
specific rotating drum boundary conditions, the small deformations
experienced by the matter explaining then the fact that this
structure of $\delta \theta(x,z) \neq 0$ can live on.

\section{Conclusion.-}

In this Letter, we have confronted a recent visco-plastic theory
for dense granular flow to numerical steady flows in rotating
drums. The existence of a univocal relation between a scalar local
friction coefficient -- defined as the ratio between the norm of
the deviatoric stress tensor to the pressure -- and the so-called
inertial number conjectured in the theory is in first
approximation fairly well verified over the {\it whole} drum, in
{\it both} the creeping ``solid'' phase and the avalanching phase
at the free surface, for inertial numbers ranging from $10^{-5}$
to $5\times10^{-2}$. Small dilatancies evidenced at the flow
surface are also found to be univocally related to the inertial
number. On the other hand, the local alignment between deviatoric
stress and strain rate tensors prescribed in the visco-plastic
theory is found to fail significantly over {\it the whole range}
of inertial number. This leads us to argue that the visco-plastic
rheology proposed in \cite{Jop06_nature} is extremely efficient to
describe quasi-unidirectional flows as those investigated in
\cite{Iordanoff04_asmejt,Dacruz05_pre,Jop06_nature} but remains
unapplicable, when applied in its non-invariant form (eq.
(\ref{rheo_2D_bis})), to highly multi-directional flows such as
those observed in rotating drum. On the other hand, it is
particularly remarkable that the Jop \textit{et al.} rheology,
when considered in its invariant form, holds for very small values
of $I$ down to $10^{-5}$ and that even if the stress and strain
rate tensor misalignment is large. This result suggests in
particular to reconsider the previous apparent failure of the
visco-plastic rheology observed for small $I$ in other flow
geometries \cite{Dacruz05_pre,Depken,Silbert01_pre} which could
have been a partially wrong conclusion. The origin of the observed
misalignment between tensors is conjectured to be the consequence
of transient dilatancy effects linked to the small deformation
experienced by the ``solid'' part of the granular matter when
considered from a Lagrangian point of view. Finally, a quadripolar
spatial structuration of the dispersion around the perfect
collapse of the data on the visco-plastic invariant rheology is
tentatively attributed to non-local effects set by boundary
conditions. In this context, stress distribution in immobile
granular packings was found to be well described by Linear
Elasticity \cite{Ovarlez03_pre}. It will then be interesting to
see whether the introduction of an additional elastic field
describing the stress distribution in an equivalent ``immobile''
packing can explain the observed discrepancy. Work in this
direction is currently under progress.

\acknowledgments

This work was supported by the Triangle de la Physique and grant
ANR TSF NT05-1-41492.

\end{document}